\documentclass[twocolumn,aps, prl, nofootinbib]{revtex4}

\usepackage{graphics}
\usepackage{graphicx}
\usepackage{dcolumn}
\usepackage{bm}
\usepackage{amsmath,amssymb,epstopdf,color,float}

\begin{document}

\title{Hidden Depths in a Black Hole: Surface Area Information Encoded in the $\bm{(r,t)}$ Sector}
\author{Charles W. Robson\footnote{email: charles.robson@tuni.fi} and Marco Ornigotti}
\affiliation{Laboratory of Photonics, Physics Unit, Tampere University, Tampere, FI-33720 Finland}
\date{\today}

\begin{abstract}

Based on an investigation into the near-horizon geometrical description of black hole spacetimes (the so-called “$(r,t)$ sector”), we find that the surface area of the event horizon of a black hole is mirrored in the area of a newly-defined surface, which naturally emerges from studying the intrinsic curvature of the $(r,t)$ sector at the horizon. We define this new, abstract surface for a range of different black holes and show that, in each case, the surface encodes event horizon information, despite its derivation relying purely on the $(r,t)$ sector of the metrical description. This is a very surprising finding as this sector is orthogonal to the sector explicitly describing the horizon geometry. Our results provide new evidence supporting the conjecture that black holes are, in some sense, fundamentally two-dimensional. As black hole entropy is known to be proportional to horizon area, a novel two-dimensional interpretation of this entropy may also be possible.

\end{abstract}

\maketitle

\paragraph{Introduction.---}Over the past few decades, the counterintuitive idea of regarding black holes as thermodynamical systems has been very fruitful and has led to several deep results \cite{Hawking_rad,Bekenstein,Bekenstein2,four_laws,Christo1,Page_review}. It was predicted by Hawking \cite{Hawking_rad} in the 1970s that black holes must emit thermal radiation (hence they are not quite ``black") and, based on the efforts of Bekenstein and Wheeler, they were also found to contribute to the total entropy of the Universe. This latter discovery leads to arguably the most interesting concept in black hole physics: a black hole has an entropy and the value of this entropy is proportional to the surface area of the event horizon \cite{Bekenstein,Bekenstein2}. This concept has led to profound ideas, such as the holographic principle \cite{Hooft,Susskind1,Bousso} and the AdS/CFT correspondence \cite{Maldacena}, and it provides deep insights into the role played by information theory in fundamental physics \cite{Susskind2,Bekenstein3,Bekenstein_channel}. Indeed, black hole thermodynamics has been thought of for several decades as a possible window into a consistent theory of quantum gravity \cite{Wald,Carlip_review}.

One of the most striking consequences of the thermodynamical approach is the realisation that almost all of the information pertaining to a black hole is contained (or encoded) in a two-dimensional sector of its geometrical description: the ``$(r,t)$ sector" \cite{Carlip1,Teitel1}. It is known that this sector is sufficient to study Hawking radiation emission, whether one uses anomalies \cite{Christen,Wilczek,Majhi}, tunnelling \cite{Wilczek2,Banerjee}, topological techniques \cite{me1}, or the standard approach enforcing the absence of conical singularities in Euclidean spacetimes \cite{Zee}.

Less well known is the fact that the entropy of a black hole relies fundamentally on the properties of its $(r,t)$ sector: after a Wick rotation to imaginary time, in fact, the entropy becomes directly proportional to the value of a topological invariant (the Euler characteristic $\chi$) of this two-dimensional sector \cite{Carlip1,Teitel1,Teitel2}, i.e. $S=\chi A/4$, where $A$ is the surface area of the event horizon. The $(r,t)$ sector, therefore, plays a fundamental role in determining the thermodynamical properties of a black hole. However, the surface area $A$ is still calculated by considering the whole four-dimensional structure of the black hole and, to the best of our knowledge, it has yet to be linked to the $(r,t)$ sector. This is understandable as the surface area of a given black hole is explicitly described by the orthogonal sector, rather than the $(r,t)$ sector \cite{Susskind2}.

In recent years, a major advance in understanding the two-dimensional nature of black hole entropy has been made using conformal field theory techniques \cite{Carlip2,Solod}. It has been shown that it is possible to derive the Bekenstein entropy by considering an algebra of surface deformations on the black hole horizon \cite{Carlip2}; in this theory, the central charge of a subalgebra is found to be proportional to the surface area $A$ of the horizon. Yet, in all cases we know of, the area $A$ is derived by integrating over the horizon geometry, i.e. by implicitly taking information from the orthogonal sector rather than the $(r,t)$ sector.

In this paper, we show that the horizon area information of a black hole is encoded entirely in the $(r,t)$ sector. To do this, we introduce a new surface, described locally by the near-horizon geometry, which effectively mirrors the event horizon, with an altered geometry but equal area: we term this new surface the \emph{lethesurface} (from the ancient Greek word \emph{lethe}, meaning ``concealment", as the surface defined in this paper is somewhat hidden due to its definition relying on a single point of the $(r,t)$ sector). We show that the concept applies for a variety of different black hole spacetimes. This is highly unexpected as the horizon area $A$ is described explicitly by the orthogonal sector. Our results show that a dimensional reduction of a black hole description leaves rich and, importantly, complete geometrical information behind. These results chime with the growing evidence (see \cite{Carlip3} for a review) that spacetime at small distances is fundamentally two-dimensional. The importance of these results for describing black hole entropy is also discussed.

\paragraph{Schwarzschild Black Hole---} \label{sec:Sec1}

The first system we analyse is the simplest: the Schwarzschild black hole. The Schwarzschild metric describes the spacetime outside of an uncharged, nonrotating black hole (or generally any compact spherical body) of mass $M$, and is given in natural units by
\begin{equation} \label{Schwarz_met}
ds^2 = -f(r)dt^2 +\frac{dr^2}{f(r)} + r^2 d\Omega^2,
\end{equation}
where $f(r)=  1-2M/r$ is the emblackening factor of a Schwarzschild black hole, and $d\Omega^2$ is the line element on the two-sphere. The intrinsic curvature of this spacetime, measured by the Ricci scalar $R$, is identically zero everywhere. However, the intrinsic curvatures of different planes cut through the spacetime are not necessarily vanishing. For example, taking a time slice at $t=0$ (any time slice is equivalent here) and then fixing the radial coordinate $r$ at the event horizon position $r_{\mathrm{H}}=2M$ leaves only the two-sphere geometry of the horizon. The Ricci scalar of this sector is then a measure of the intrinsic curvature of a spherical surface of radius $r_{\mathrm{H}}$, given by $R^{(\theta , \phi)}\big\rvert_{r_{\mathrm{H}}}=2/r_{\mathrm{H}}^2$.

A more interesting spacetime slice is found instead by fixing the angular coordinates $\theta$ and $\phi$. This leaves the line element of the so-called $(r,t)$ sector, i.e.
\begin{equation} \label{eq:2d_Schwarz}
ds^2 = - \left( 1-\frac{2M}{r} \right)dt^2 + \left( 1-\frac{2M}{r} \right)^{-1}dr^2,
\end{equation}
which describes the spacetime curvature experienced by a radially-moving observer outside of the black hole. More generally, this sector is what is ``felt" by fields in the vicinity of the Schwarzschild black hole horizon at any angular coordinate values \cite{Murata}. The Ricci scalar of this sector is $R^{(r,t)}=4M/r^3$. Setting the radial coordinate to $r=r_{\mathrm{H}}$ then yields the intrinsic curvature of the $(r,t)$ sector at the horizon. After substituting in the value $M=r_{\mathrm{H}}/2$, the intrinsic curvature of this sector is found to be
\begin{equation} \label{Schwarz_Ricci}
R^{(t,r)}\big\rvert_{r_{\mathrm{H}}} = \frac{2}{r_{\mathrm{H}}^2},
\end{equation}
which mirrors the intrinsic curvature of the orthogonal, purely spatial sector, derived above, at the horizon.

\begin{figure}
\centering
\includegraphics[width=0.2\textwidth]{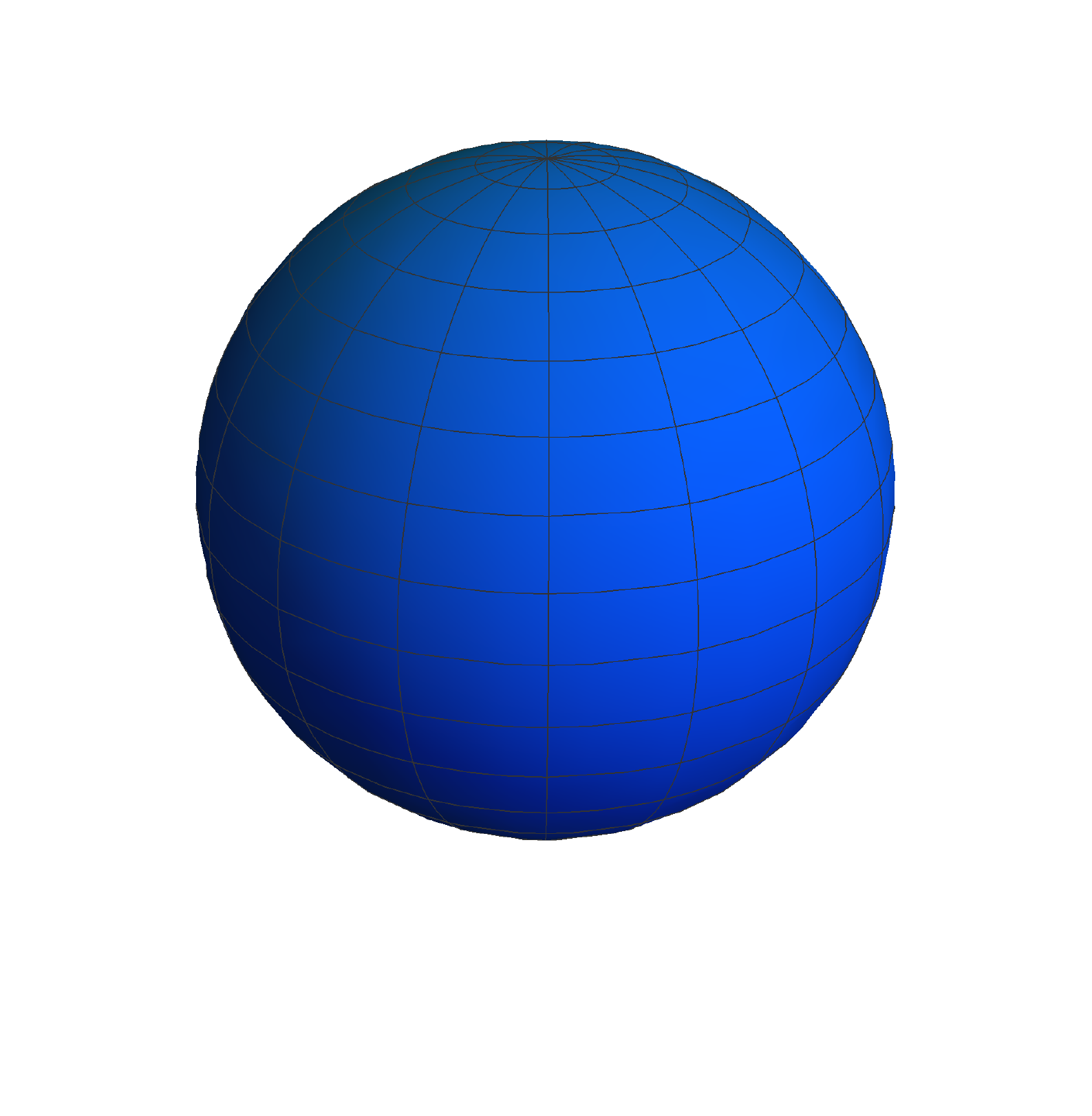}
\caption{The Schwarzschild \emph{lethesurface}: the surface of a sphere. A sphere's surface has constant, positive intrinsic curvature.}
\label{figure1}
\end{figure}

Eq. \eqref{Schwarz_Ricci} signifies that the intrinsic curvature at the horizon of the $(r,t)$ sector is the same as that at any point on the surface of a sphere of radius $r_{\mathrm{H}}$. We extrapolate from (\ref{Schwarz_Ricci}) by saying that it describes a point on a new, abstract surface; the most natural, symmetric choice being a surface having the constant curvature (\ref{Schwarz_Ricci}) everywhere, i.e. a sphere of area $A=4\pi r_{\mathrm{H}}^2$. This is an indication that the $(r,t)$ sector contains the information required to determine the event horizon area. Equation \eqref{Schwarz_Ricci} implies that the geometry at the horizon describes the local region of a \emph{hidden} surface, that we name the \emph{lethesurface}. The presence of this surface is only revealed once the geometry of the $(r,t)$ sector at the horizon is studied in detail. The \emph{lethesurface} for a Schwarzschild black hole is then simply a two-sphere with an area equal to that of the event horizon itself (as shown in Fig. \ref{figure1}). We argue that the horizon area $A$ is encoded in the $(r,t)$ sector, despite the sector not explicitly describing the horizon geometry as well as the fact that one of its dimensions is temporal. This is the first result of our paper.

At first glance, this result may appear trivial due to the spherical symmetry of the Schwarzschild spacetime. However, note that this mirroring of the $(\theta,\phi)$ sector Ricci scalar does not apply for a general spherically symmetric compact object. For example, using the Schwarzschild metric (\ref{Schwarz_met}) to describe the vacuum spacetime exterior to a star, one can see that the Ricci scalars of the two sectors at the surface of the star never match -- its radius would have to decrease to the Schwarzschild radius in order for this to occur: something impossible without the star itself becoming a black hole.

To make our argument more convincing, and to show that it is possible to define a \emph{lethesurface} for a more complicated black hole having less symmetry, in the remainder of this paper we discuss the properties of \emph{lethesurfaces} for both charged and rotating black holes. In particular, we will show that each black hole naturally has associated to it a \emph{lethesurface} (in some cases, more than one) and that, strikingly, in general, the \emph{lethesurface} and the event horizon have vastly different geometries yet always equal area. 

\paragraph{Charged Black Hole---} \label{sec:Sec2}

The spacetime exterior to a black hole with electric charge $Q$ can be described by the metric in Eq. \eqref{Schwarz_met} with the emblackening factor
\begin{equation}
f(r)= \left( 1-\frac{2M}{r} + \frac{Q^2}{r^2} \right);
\end{equation}
this is the Reissner-Nordstr\"{o}m metric. Dimensionally reducing the metric as before gives the $(r,t)$ sector of a charged black hole, namely
\begin{equation}
ds^2 = - \left( 1-\frac{2M}{r} + \frac{Q^2}{r^2} \right)dt^2 + \left( 1-\frac{2M}{r} + \frac{Q^2}{r^2} \right)^{-1}dr^2,
\end{equation}
with Ricci scalar at the horizon given by
\begin{equation} \label{eq:RN_Ricci}
R^{(t,r)}\big\rvert_{r_{+}} = \frac{2}{r_{+}^2} - \frac{4Q^2}{r_{+}^4},
\end{equation}
where $r_{+}=M+\sqrt{M^2-Q^2}$ is the outer horizon of the Reissner-Nordstr\"{o}m black hole \cite{Carroll1}.

\begin{figure}
\centering
\includegraphics[width=0.2\textwidth]{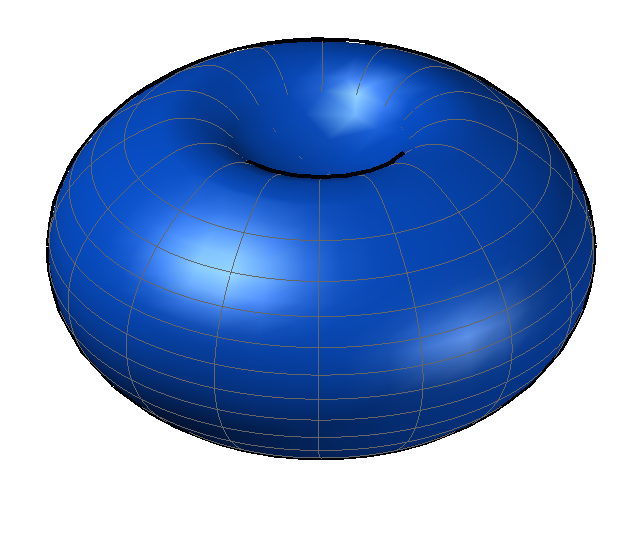}
\caption{Artist's impression of a \emph{lethesurface} for a Reissner-Nordstr\"{o}m black hole, for an arbitrary charge value in the range $2\sqrt{2}M/3<Q \leq M$. The image is intended only to represent the intrinsic curvature of the surface. The intrinsic curvature is positive near the equator; the dip shown at the pole illustrates the intrinsic curvature becoming flat and then negative as one moves towards either pole. An artist's rendering is necessary as the surface cannot be globally embedded in Euclidean 3-space.}
\label{figure2}
\end{figure}

The local geometry described by Eq. \eqref{eq:RN_Ricci} is very interesting, in particular due to the unusual form of the Ricci scalar. To see this, let us first consider the two limiting cases of vanishing charge and extremal black hole. In the uncharged limit, i.e. $Q\rightarrow 0$, the Ricci scalar above is again that of a point on the surface of a sphere of area equal to the Schwarzschild horizon area, as one would expect. In the maximally charged (extremal) limit $Q \rightarrow M$, the geometry locally describes a pseudosphere, with a surface area equal to that of the extremal charged black hole's horizon.

For the intermediate case $Q < M$, on the other hand, the intrinsic curvature given by Eq. \eqref{eq:RN_Ricci} has two competing curvature terms, one positive and one negative. Depending on the magnitude of the charge $Q$, the total curvature can be positive, negative, or zero. Clearly, this geometry is highly nontrivial. Remarkably, this curvature again naturally corresponds locally to a surface with the same area as the event horizon. The global geometry of this \emph{lethesurface} can be described by the line element
\begin{equation} \label{RN_lethesurface}
ds^2 =g(\alpha,Q)d\alpha^2 + \frac{r_{+}^4}{g(\alpha,Q)}\mathrm{sin}^2 \alpha d\beta^2,
\end{equation}
where $\alpha\in[0,\pi]$, $\beta\in[0,2\pi)$, and $g(\alpha,Q)= \left[ r_{+}^2 - \left( \frac{Q^2}{2} \right) \mathrm{sin}^2 \alpha \right]$.

Note that there is no possibility of a \emph{lethesurface} being defined for the charged black hole with such a high degree of symmetry as in the Schwarzschild case treated above, without the surface behaving pathologically as the parameter $Q$ evolves. For example, a spherically symmetric \emph{lethesurface} transitioning from positive to zero curvature as the charge increased to $Q=2\sqrt{2}M/3$ would have to be a sphere growing to infinite radius (and, of course, infinite area); continuing to enforce a high degree of symmetry as the charge increased further, the infinitely large sphere would then be forced to discontinuously transition to a surface with constant negative curvature. By far the most natural choice of Reissner-Nordstr\"{o}m \emph{lethesurface} is then that described by (\ref{RN_lethesurface}), without any discontinuous transitions as the value of $Q$ evolves. (As discussed later on, the surface (\ref{RN_lethesurface}) is in fact familiar from black hole theory, as it has the same geometric structure as the outer event horizon of a Kerr black hole \cite{Visser}, as can be proven through a simple reparameterisation.)

The line element in Eq. \eqref{RN_lethesurface} in fact describes a family of geometries which, depending on the magnitude of $Q$, covers the surface of a sphere ($Q=0$), the surface of a spheroid ($0<Q<2\sqrt{2}M/3$), or, for $2\sqrt{2}M/3<Q \leq M$, an intermediate shape resembling a spheroid near the equator, but with local negative curvature at the poles. The \emph{lethesurface} for a charged black hole is visualised in Fig. \ref{figure2}. The area of all these surfaces can be directly calculated from Eq. (\ref{RN_lethesurface}) and is given by $A=4\pi r_{+}^2$. Moreover, the Ricci scalar for the surface described by Eq. \eqref{RN_lethesurface} at either pole is equivalent to Eq. \eqref{eq:RN_Ricci}. Despite its unusual shape, the area of the \emph{lethesurface} exactly corresponds to the horizon area of the charged black hole for any physical value of Q.

We conclude our analysis of charged black holes by noticing two puzzling features. The first concerns the curvature described by the Ricci scalar in Eq. \eqref{eq:RN_Ricci}. When the black hole charge reaches the special value $|Q|=2\sqrt{2}M/3$ (which is below the extremal value $|Q|=M$), the sign of the Ricci scalar changes from positive to negative. It is not clear whether this change of sign occurring at $|Q|=2\sqrt{2}M/3$ corresponds to a physically important transition. The second feature concerns a seemingly hidden connection between charged and rotating black holes: the \emph{lethesurface} of a charged black hole, described by Eq. \eqref{RN_lethesurface}, has the same geometrical structure as the event horizon of a Kerr black hole \cite{Smarr}, as noted earlier. This unexpected link is compelling and should be investigated further.

\begin{figure}
\centering
\includegraphics[width=0.2\textwidth]{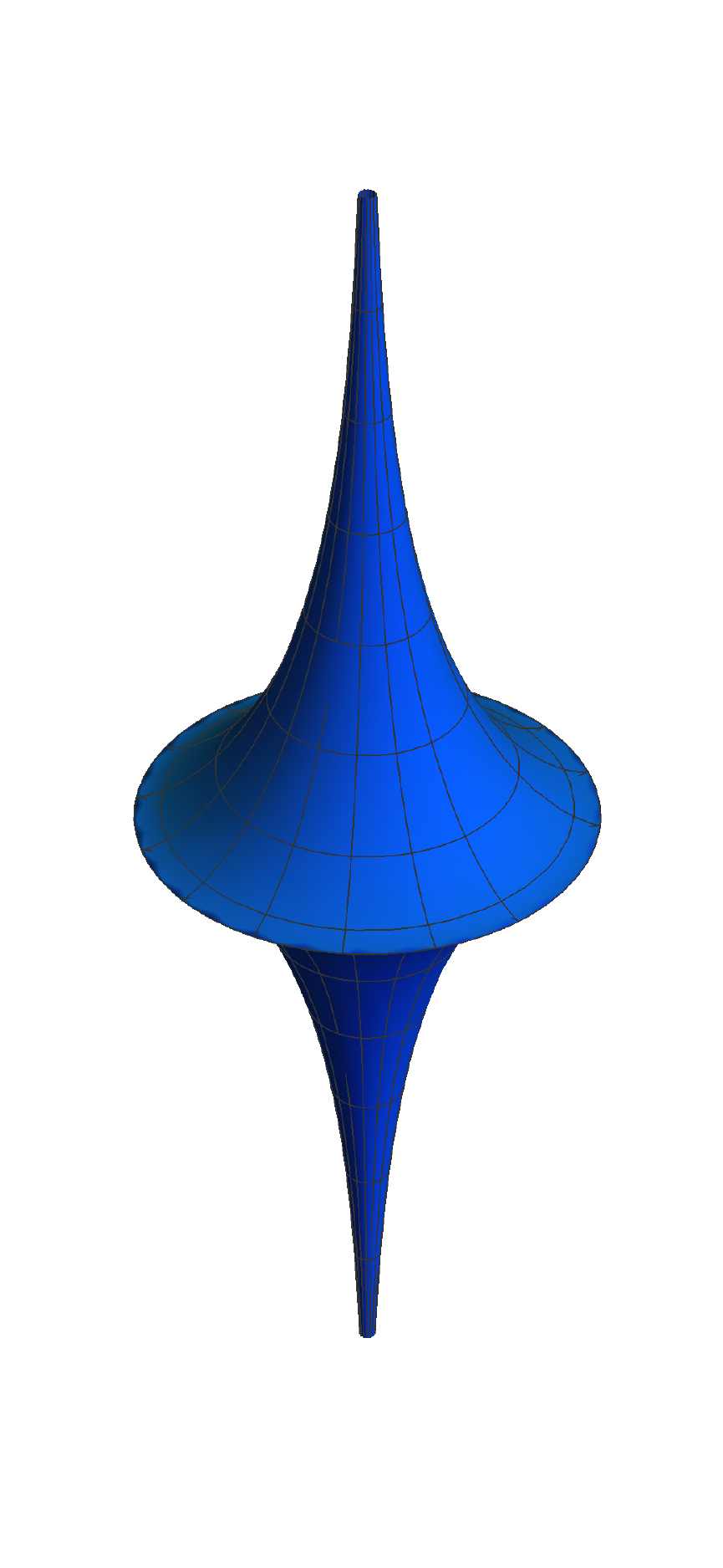}
\caption{A Kerr \emph{lethesurface}: the pseudosphere. This is a hyperbolic surface formed by rotating a tractrix around its asymptote \cite{Bonahon} -- despite its infinite extent, a pseudosphere of radius $L$ (with Ricci scalar $R=-2/L^2$) has a finite surface area of $4\pi L^2$ \cite{Needham}.}
\label{figure3}
\end{figure}

\paragraph{Kerr Black Hole---} \label{sec:Sec3}

The Kerr black hole rotates around its axis with angular momentum per unit mass $a$, the rotation flattening its event horizon towards a spheroidal shape \cite{Carroll1,Kerr1}. Due to a lack of spherical symmetry, a useful geometrical description of this spacetime in terms of only radial and temporal coordinates would seem unlikely but, strikingly, a two-dimensional effective metrical description, valid near the event horizon $r_{+}=M+\sqrt{M^2-a^2}$, has been found: the effective metric of the $(r,t)$ sector is given by \cite{Murata}
\begin{equation} \label{eq:Soda}
ds^2 = - K(r) dt^2 + \frac{1}{K(r)}dr^2,
\end{equation}
with $K(r)=(r^2 - 2Mr + a^2)/(r_{+}^2 + a^2)$. The intrinsic curvature of this lower-dimensional sector provides complete information concerning the surface area of the horizon, up to the extremal value of $a$. The Ricci scalar for the geometry defined above is
\begin{equation} \label{eq:Kerr_Ricci}
R^{(t,r)} = -\frac{2}{r_{+}^2 + a^2}.
\end{equation}
(Notice that, contrary to the other cases, the Ricci scalar for the effective $(r,t)$ sector of a Kerr black hole need not be evaluated at the event horizon position $r_+$, since the effective metric given above is by construction only defined near the horizon \cite{Murata}.)

Eq. \eqref{eq:Kerr_Ricci} describes an intrinsic curvature identical to that at a point on the surface of a pseudosphere of radius $\left( r_{+}^2 + a^2 \right)^{1/2}$. The area of the \emph{lethesurface} is then the area of such a pseudosphere, i.e. $A=4\pi\left( r_{+}^2 + a^2 \right)$. This is precisely the Kerr black hole's outer event horizon area. For a rotating black hole, therefore, one natural choice of \emph{lethesurface}, defined using an effective near-horizon geometry, is a pseudosphere, a shape quite unlike that of the event horizon -- see Fig. \ref{figure3}.

The horizon of a Kerr black hole can, in fact, be hyperbolic near the poles (and only near the poles) if the rotation is rapid enough. However, the horizon never becomes hyperbolic over its entire surface in any limit \cite{Smarr}. This is another indication that, in general, the \emph{lethesurface} and event horizon of a given black hole have very different geometries yet equal areas.

\paragraph{Uniqueness of the lethesurface---} \label{sec:Sec4}

There is an important point to make concerning the uniqueness of the \emph{lethesurface} for a given black hole. To do this, we consider a different dimensional reduction of the Kerr black hole, using the so-called Boyer-Lindquist coordinates \cite{Carroll1}. This reduction can be performed by fixing the $\theta$ variable at either pole ($\theta=0$ or $\pi$). Notably, the Ricci scalar of the $(r,t)$ sector for this new geometry at the horizon again locally and naturally describes a surface with an area matching that of the event horizon, but in this case not a pseudosphere \cite{Note1}. Furthermore, the curvature described by Eq. \eqref{eq:Kerr_Ricci} can also naturally describe the local geometry of a \emph{twisted} pseudosphere (also known as Dini's surface \cite{Encyclo}) with a much more complicated structure. Nevertheless, the form of this twisted pseudosphere (though not uniquely determined by  the Ricci scalar) can be easily constructed such that its surface area is equal to that of the Kerr horizon, by the tuning of free parameters.

One must take care, therefore, to define the \emph{lethesurface} concept unambiguously. Whether there can be a strict one-to-one mapping between event horizon and \emph{lethesurface} for a given black hole -- a uniqueness theorem -- is currently unknown. We conjecture that it should be possible to at least construct a theorem whereby one could tightly constraint the choice of \emph{lethesurface} for a given black hole spacetime. There are two pieces of evidence supporting this conjecture. One is that it would seem, based on our results, that one should always construct the surface having the most symmetry; this favouring of surfaces endowed with the greatest symmetries adds a significant constraint. The second, stronger piece of evidence is that the \emph{lethesurfaces} constructed so far (except for the pseudosphere in Fig. \ref{figure3} which originates from an effective geometry) belong to a special geometric class in that they are all minimal surfaces in an appropriate ambient space: i.e. each surface has a vanishing mean extrinsic curvature or, equivalently, has the minimum surface area of all surfaces having the same boundary \cite{Gray_book}. The Schwarzschild \emph{lethesurface} (shown in Fig. \ref{figure1}) is a minimal surface in a $t=$ const. hypersurface of Schwarzschild spacetime. The surface shown in Fig. \ref{figure2}, and that of the Kerr black hole found using Boyer-Lindquist coordinates as alluded to earlier, are also minimal surfaces in Kerr spacetime on a $t=$ const. hypersurface (this is the most natural ambient spacetime as both \emph{lethesurfaces} have the geometry of a Kerr black hole's outer event horizon).

The role of minimal surface theory and that of symmetry may be crucial in forming a uniqueness theorem for the duality presented in this paper and efforts are currently underway to rigorously formulate it. (For further reading on the links between minimal surfaces and black hole theory, see the excellent review on entanglement entropy of black holes by Solodukhin \cite{Solod2}.)

The main result of our paper is that we have shown that it is possible to find the horizon area of an uncharged, charged, and rotating black hole, using only information contained in the $(r,t)$ sector. Importantly, our results are valid for any physical value of system parameter, up to and including extremal values, and are strongly constrained by considerations of symmetry and area minimisation. 
\paragraph{Extremal Kerr-Newman Black Holes---} \label{sec:Sec5}
An immediate application of our results is that they explain a previously mysterious feature of extremal black holes. Extremal Kerr-Newman black hole spacetimes on the polar axis (i.e. at $\theta=0$ or $\pi$) are known to be $\mathrm{AdS}_{2}$. As noted by Bardeen and Horowitz in Ref. \cite{Horowitz}: ``the area of the [extremal] event horizon is related to the effective cosmological constant of $\mathrm{AdS}_{2}$ in a universal way that is independent of whether the extreme black hole has only angular momentum, charge, or both". The authors \cite{Horowitz} then speculate that this feature may play a role in the context of the AdS/CFT correspondence \cite{Maldacena}. 

The explanation for this ``universal" relation between the effective cosmological constant of $\mathrm{AdS}_{2}$ spacetime and the extremal horizon area follows directly from our results.

We first note that $\mathrm{AdS}_{2}$ space is pseudospherical and that the AdS length scale $L$ is just the radius of said pseudosphere. Now consider the fact that, near the horizon, and on the polar axis, the $(r,t)$ sector of the extremal Kerr-Newman black hole locally describes a pseudosphere, i.e. the extremal Kerr-Newman black hole has a pseudospherical \emph{lethesurface}.  The area of this \emph{lethesurface} is then the area of a pseudosphere of radius $r_{0}$, i.e. $4\pi r_{0}^2$ (using the notation of Ref. \cite{Horowitz}). Horowitz and Bardeen's interesting ``universal" relation can be then solely explained in terms of classical general relativity, using the concept of \emph{lethesurface}.
\paragraph{Lethesurfaces and Entropy---} \label{sec:Sec6}
The entropy of a black hole can be put in relation to its $(r,t)$ sector (via its topology) by the relation $S=\chi A/4$. In this formula, however, the area $A$ is the area of the event horizon. We can modify the entropy law as follows: $S=\chi A_{ls}/4$, i.e. the entropy is proportional to the area of a \emph{lethesurface} ($A_{ls}$) associated to the black hole. As has been articulated by Susskind, ``information equals area" for a black hole \cite{Susskind3}. This ``area" of course refers to that of the event horizon. Motivated by our results, perhaps the definition of ``area" in Susskind's equation can be generalised such that it holds not only for horizon area, but also for \emph{lethesurface} area. This is the second important result of our paper: by introducing the concept of a \emph{lethesurface} for a black hole, we can now completely determine its entropy purely by probing the $(r,t)$ sector, without the need to access information from the rest of the black hole metric.

For black hole entropy to be associated to our newly-defined surface in a meaningful way, it would seem to be required that the microscopic states responsible for this entropy must be localised on this surface somehow. Recent results from string theory suggest that it is precisely at the point $r=r_{\mathrm{H}}$ (the so-called ``tip of the cigar", meaning the origin of the Euclidean Schwarzschild $(r,\tau)$ sector), that certain zero modes are localised -- zero modes which have been argued to account for the entropy \cite{String}. As we have shown, it is precisely this point $r=r_{\mathrm{H}}$ of the $(r,t)$ sector that is special when deriving our geometrical results. It would be interesting to see whether there are any links to be made between our work and these string-theoretic considerations.

Finally, it may be fruitful to investigate whether any relations exist between our (1+1)-dimensional approach and recent research on the (1+1)-dimensional nature of black holes using information theory \cite{Bekenstein_channel} and dimensional analysis \cite{Hebei}.

\paragraph{Conclusions.---}

In this paper, we have introduced the concept of the \emph{lethesurface}, i.e. an abstract surface associated to each black hole, which is shown to contain geometrical information concerning the event horizon, despite originating from an orthogonal metric sector. This is a completely new feature in black hole theory and is a direct result of only classical general relativistic considerations. By investigating the $(r,t)$ sector of each black hole spacetime, this previously-undefined surface is found to have equal surface area to the corresponding black hole's horizon for uncharged, charged, and rotating black holes, for any parameter value up to and including extremality. A formerly unexplained feature of extremal Kerr-Newman spacetimes, namely a link between horizon area and an effective $\mathrm{AdS}_{2}$ cosmological constant, is fully explained by our results. The encoding of surface area information, and therefore entropy, at the horizon of the $(r,t)$ sector as established in this paper adds strong evidence in favour of the intrinsic (1+1)-dimensional nature of black holes and perhaps spacetime in general. An extension of the results presented in this paper to those spacetimes with nonzero cosmological constant, as well as further probing of the role of information theory, will be the subject of future work.

\paragraph{Acknowledgments.---}
The authors wish to thank Dr. Fabio Biancalana and Zoe Wyatt for fruitful discussions and to thank Kexin Wang for his help with the figures.

\end{document}